\documentclass[preprint,5p,times]{elsarticle}

\usepackage[T1]{fontenc}
\usepackage{gensymb}
\usepackage{graphicx}
\usepackage{tabularx}
\usepackage{footnote}
\usepackage{xspace}
\usepackage{amsmath}
\usepackage{natbib}

\usepackage{mathrsfs}
\usepackage{dcolumn}
\newcolumntype{d}{D{.}{.}{1} }

\usepackage{lineno}
\usepackage{xcolor}

\newcolumntype{Y}{>{\centering\arraybackslash}X}

\newcommand{\isotope}[2]{$^{#1}$\text{#2}}

\usepackage[unicode=true,bookmarks=true,bookmarksnumbered=false,bookmarksopen=false,breaklinks=false,pdfborder={0 0 1},backref=false,colorlinks=true]{hyperref}
\hypersetup{urlcolor=blue, citecolor=blue}

\begin{document}

\title{Offline Commissioning of the St. Benedict Gas Catcher}

\author[nd]{F. Rivero}
\author[ens]{D. Guillet}
\author[nd]{M. Brodeur \corref{cor1}}
\author[anl]{J.A. Clark}
\author[nd]{A.M. Houff}
\author[nd]{J.J. Kolata}
\author[nd]{B. Liu}
\author[nd]{J. McRae}
\author[nd]{P.D. O'Malley}
\author[nd]{W.S. Porter}

\author[nd]{C. Quick\fnref{fn1}}
\fntext[fn1]{Present address: Department  of Physics and Astronomy, University of Tennessee, Knoxville, TN 37996, USA}

\author[anl,uc]{G. Savard}
\author[anl]{A.A. Valverde}
\author[nd]{R. Zite}

\address[nd]{Department of Physics and Astronomy, University of Notre Dame, Notre Dame, Indiana 46556, USA}
\address[ens]{Département de Physique, École normale supérieure Paris-Saclay, 4 avenue des Sciences, 91190 Gif-sur-Yvette, France}
\address[anl]{Physics Division, Argonne National Laboratory, Lemont, Illinois 60439, USA}
\address[uc]{Department of Physics, University of Chicago, Chicago, IL 60637, USA}
\address[utk]{Department of Physics and Astronomy, University of Tennessee, Knoxville, Tennessee 37996, USA}

\cortext[cor1]{mbrodeur@nd.edu} 

\begin{abstract}
Precision measurements of $\beta$ decay transitions offer a promising channel through which the Standard Model (SM) can be probed. There is currently an ongoing effort to increase the precision on measurements of $\mathcal{F}t$-values for superallowed $\beta$ decay transitions between mirror nuclides. These allow for a determination of $V_{ud}$ which is complementary to that obtained from pure Fermi $0^+ \rightarrow 0^+$ transitions. The Superallowed Transition BEta-NEutrino Decay Ion Coincidence Trap (St. Benedict), under construction at the Nuclear Science Laboratory (NSL) at the University of Notre Dame, seeks to measure the Fermi-to-Gamow-Teller mixing ratio for transitions between mirror nuclei in order to expand the list of  nuclides from which $V_{ud}$ can be extracted. Production and selection of the species of interest will be done in-flight, using the \textit{TwinSol} magnetic separator system. The first element of St. Benedict will be a large volume gas catcher which will thermalize radioactive ion beams for low energy delivery to the rest of the system. Offline commissioning of this gas catcher has been completed using an internal potassium source, and the device demonstrated a transport efficiency upwards of 95\%  for pressures of 66 mbar and lower. 
\end{abstract}

\maketitle


\section{Introduction}

The Standard Model (SM) \cite{Group2020} stands as the most robust framework through which we understand the fundamental building blocks of the universe. However Beyond the Standard Model (BSM) phenomenon such as matter-antimatter asymmetry, dark matter, and dark energy \cite{Lagouri2022} make it clear that this picture is incomplete. Low energy studies of nuclear beta-decay offer a promising probe into the electro-weak interaction \cite{Falkowski2021, Naviliat2016} which is complimentary to high-energy searches. 

The unitarity of the Cabibbo-Kobayashi-Maskawa quark mixing matrix \cite{Group2020}, which describes a rotation between quark eigenstates, is one of the most critical tests of the SM. There are many mathematical tests of unitarity which can be performed, however the one that currently provides the most precise value \cite{Hardy2014, Hardy2018} is the, so-called, top row test
\begin{equation}
    \sum_i |V_{ui}|^2 = |V_{ud}|^2 + |V_{us}|^2 + |V_{ub}|^2 = 1
\end{equation}
which considers only the coupling of the up quark to the negatively charged quarks. 

In recent years the $V_{ud}$ element of this matrix has garnered considerable attention, after a revisiting of theoretical corrections yielded a value for the term which brought the top row test to a value $\sim 3\sigma$ below unitarity \cite{Hayen2020vud, Czarnecki2019}. This has spurred a flurry of effort on both theoretical and experimental fronts to improve accuracy and increase precision of the accepted value of $V_{ud}$ \cite{Hardy2014, Hardy2018}. 

One such effort consists of extracting $V_{ud}$ from superallowed transitions between mirror nuclei. As a whole, this ensemble of measurements currently lacks a competitive precision when compared to $0^+ \rightarrow 0^+$ decays. This is due to the added need to determine the Fermi-to-Gamow-Teller mixing ratio, $\rho$, for such transitions.
 
The Superallowed Transition BEta-NEutrino Decay Ion Coincidence Trap (St. Benedict), being constructed in the Nuclear Science Laboratory (NSL) at the University of Notre Dame, will fulfill this need by determining $\rho$ for a number of mixed-mirror beta decays. The species of interest will be produced using the \textit{TwinSol} magnetic recoil separator \cite{Omalley2022}, which consists of two superconducting solenoid magnets that filter out nuclear species based on their magnetic rigidity. From there the radioactive ion beam (RIB) will pass through a dipole bending magnet into St. Benedict. 

St. Benedict \cite{Omalley2020, Brodeur2023} will consist of a large volume gas catcher, to stop the fast RIBs, a differentially pumped region, which includes a radio-frequency (RF) carpet \cite{Davis2022comish, Davis2022trans}, a radio-frequency quadrupole (RFQ) ion guide \cite{Zite2026}, a RFQ buncher to cool and bunch the beam \cite{Burdette2025}, and a measurement Paul trap, where the final decays can be observed \cite{Porter2025}. 

This article presents the off-line commissioning of the St. Benedict gas catcher prior to its on-line installation. The goal of the commissioning was two-fold: assess the transport efficiency of thermalized ions at various pressures; and obtain operational parameters at those pressures. This allows for the study of ion transport separate from stopping, while also providing a set of operational settings once on-line operation commences.

\section{Operational principles}
Gas catchers, also sometimes called gas stoppers, are, at their most basic level, high pressure gas cells used to stop fast-moving ion beams for low-energy extraction. They have proven to be a powerful tool in the study of short lived radionuclides \cite{Arje1986, Savard2014, Scielzo2012} since they allow for precise delivery of species produced in-flight to low-energy beam experiments such as ion-traps and laser spectroscopy. 

For St. Benedict, the species of interest will be nuclei produced as a RIB with energies in the range of 10-40 MeV. In order to operate at the pressures necessary for efficient transport, the majority of the beam's energy must be removed before it enters the actual gas volume. This is done with thin, solid films or foils of a suitable material such as aluminum or Mylar. These are typically employed as a combination of a variable-angle degrader and the gas-cell entrance window, which separates the high-vacuum upstream beam-line from the gas region. Once inside the gas, beam energies are typically on the order of 1-6 MeV. This remaining energy is lost via elastic collisions with the gas atoms while retaining their charge \cite{Savard2011}. Noble gasses are the most suitable choices for stopping gas, given that they are chemically inert and have large ionization potentials. This gas catcher operates using ultra high purity helium. 

The thermalized RIB can then be manipulated with a combination of RF and DC electric fields which are supplied by a series of electrodes lining the interior of the walls. The speed at which they can move through this medium is given by 
\begin{equation}
    v_{ion} = K_0 \cdot E \cdot \frac{P_0 T}{P T_0}
\end{equation}
where $K_0$ is the reduced ion mobility, $E$ is the drag field strength, $P$ and $T$ are pressure and temperature, respectively, and $P_0$= 1000 mbar $T_0$ = 273 K \cite{Savard2011}. For reference, the reduced ion mobility for \isotope{39}{K}$^+$ in room temperature helium is around 21 cm$^2$/V$\cdot$s \cite{Viehland2012}. 

The electrical connections are such that adjacent electrodes receive alternating RF signals 180$^\circ$ out of phase. A resistor chain connects electrodes of the same RF phase, and a constant voltage is applied across this chain to create the drag field that pushes ions towards the exit of the chamber. Once the ions are within the vicinity of the exit nozzle, the bulk gas-flow is able to extract ions from the gas cell \cite{Savard2011}. 

One of the most critical features of gas stopping cells is that they should adhere to UHV standards to minimize the presence of contaminants. Contaminants can become ionized by the incoming beam and charge exchange with the thermal ions of interest or ionized helium. In turn this can lead to beam loss through recombination, or the formation of molecules, both neutral and charged. 

Besides the incoming beam energy, the primary tuning parameters for adjustment of the stopping  and extraction profile in a gas catcher are the DC drag field strength, the RF amplitude, and the gas pressure. The drag field strength will largely dictate the extraction time, the RF amplitude determines the radial confinement, and the gas pressure affects the stopping power, beam scatter, electric field breakdown threshold, and efficacy of voltages in providing the necessary field strengths for both confinement and transport. 

The RF amplitude, for purposes of radial confinement, needs only to reach a certain threshold for effective field strength in order to keep all ions away from the interior walls. Likewise the DC field only needs to be strong enough to bring the ions to the vicinity of the exit nozzle within an acceptable time-scale at a given pressure. This is often guided by the lifetime of the radioactive species. The most significant operational parameter is the gas pressure as it will also influence the effect of the other two parameters. 

The most direct effect of an increase in gas pressure is an increase in the stopping power of the device. However, a secondary effect is the significant slowing of ion transport due to a larger number of collisions with the ambient gas. Increased gas pressure also requires much larger RF amplitudes to provide the same quality of containment, in such a way that the necessary voltages begin to approach the breakdown field strength, as dictated by the Paschen curve \cite{Paschen1889}. For the pressures at which a gas catcher typically operates (40-130 mbar), the breakdown field in helium is on the order of $\sim$800-1100 of V/cm \cite{Das2018}. Typical electrode spacing in these gas cells is on the order of 1 mm. This limits the maximum voltage difference between RF phases to $\sim$80-100 V. For DC voltages, the most significant concern is sparking between the highest-potential electrode and the chamber  wall. As such, it is advisable to construct these devices in such a way as to allow for a voltage to be applied to the chamber walls in order to minimize the potential difference between the electrodes and the wall.

\section{Gas Catcher Characteristics}

\begin{figure*}
    \centering
    \includegraphics[width=\linewidth]{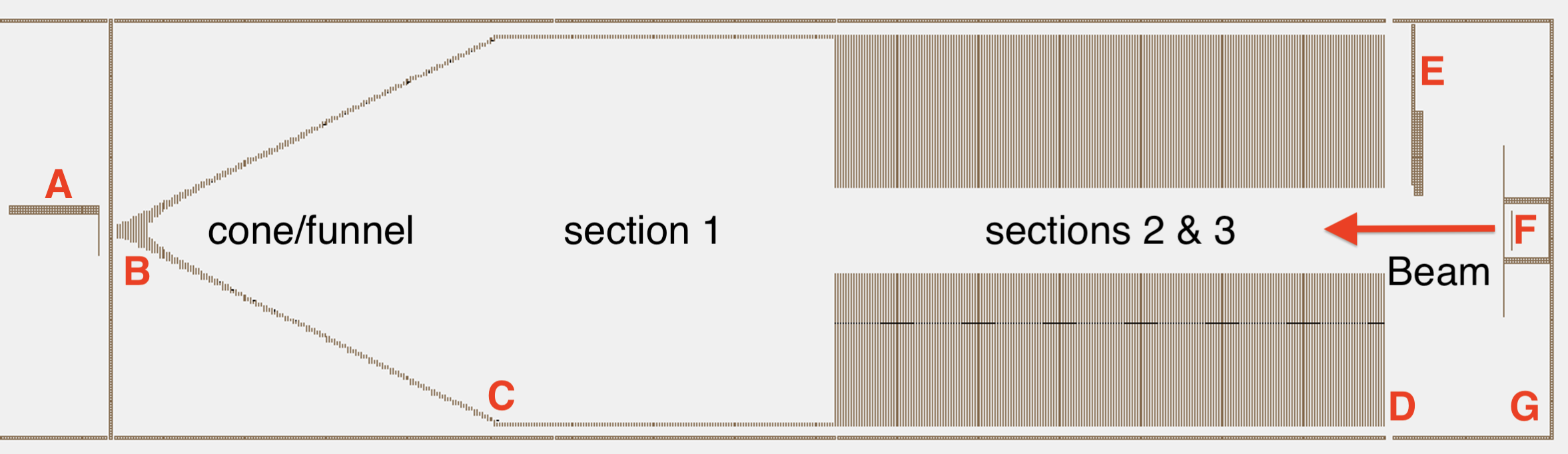}
    \caption{Cross sectional diagram of the gas catcher during offline commissioning with all of the independently tunable electrodes labeled. (A) is FC2, (B) is the location of the ``cone low'' electrode, (C) marks the end of the body electrodes and beginning of the cone, where ``body low'' and ``cone high'' sit next to each other, (D) marks the location of the ``body high'' electrode, (E) is FC1, shown in its retracted position, (F) is the ion source and square anode plate, and (G) is the window. The thicker electrodes in section 2 and 3 are the spokes on those electrodes. The ``nozzle'' electrode is represented here as the vertical line between the cone and FC2. This image was generated with the SIMION program.}
    \label{fig:GCDiagram}
\end{figure*}

\begin{figure}
    \centering
    \includegraphics[width =\linewidth]{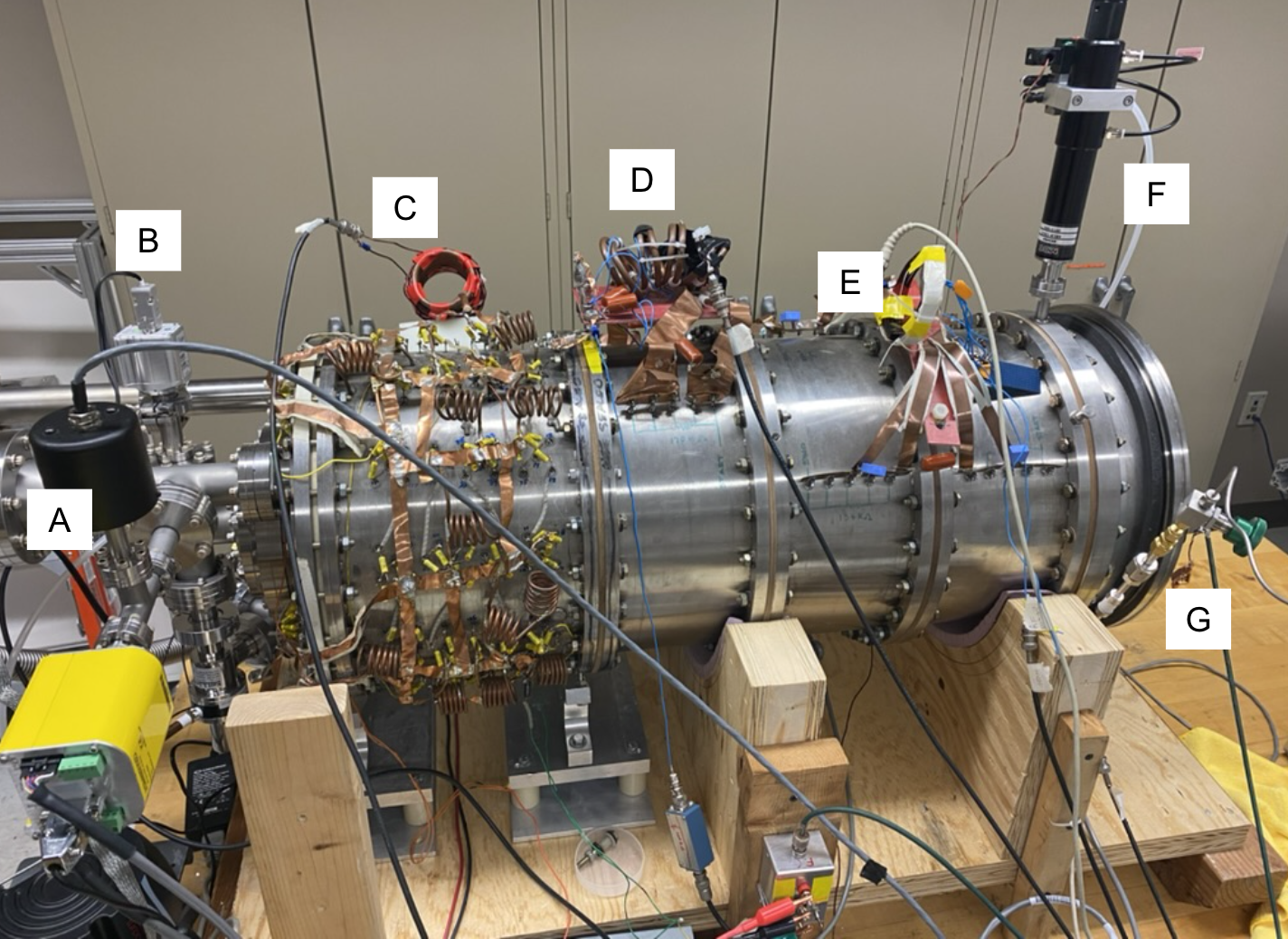}
    \caption{Photo of the gas catcher during offline commissioning. Beam direction is from right to left. (A) is an MKS type 121A Absolute Baratron Manometer mounted to measure the pressure in the main volume of the chamber. (B) is a Piezogauge mounted on the exit side of the nozzle. (C-E) are the transformers which supply RF power to the cone (C), section 1 (D), and sections 2/3(E) of the electrodes. (F) is the linear drive used to move FC1 in and out of the path of the ions. (G) is the insulated VCR-O gas line used to supply helium.}
    \label{fig:GC_setup}
\end{figure}

The St. Benedict gas catcher was formerly used, and subsequently de-commissioned, at Argonne National Laboratory's ATLAS facility. As such, a number of minor repairs, primarily to electrical connections, and replacement of electrical components, had to be performed once it was received at the NSL. Once repairs were complete, the chamber was sealed and pumped down to a pressure of $1.3\times10^{-8}$ torr in approximately 8 days, using a 500 L/s turbo pump. Once the chamber reached a satisfactory base pressure, it was leak-checked using an SRS RGA100 residual gas analyzer. Figure \ref{fig:GC_setup} shows the gas catcher as it was during the commissioning process, and indicates several of the external elements of the system. 

\subsection{Physical Dimensions}
The primary volume of the gas catcher measures roughly 33" (84 cm) in length, and 12.6" (32 cm) in diameter. The chamber is divided into four sections which will henceforth be referred to, in order from entrance to exit, as sections 3, 2, 1, and cone. Sections 1-3 are collectively referred to as the ``body'' of the chamber, while the cone is simply referred to as such. At the exterior, each of the body sections are 6" long, while the cone section is 9" long. Each of these sections is separated by a 0.5" thick insulating centering ring and a custom, spring-loaded gasket. 

Interior electrodes are 0.9 mm thick, with 0.635 mm of space between them. The electrodes in section 1 and the cone are a simple ring shape. The electrodes in sections 2 and 3 are similar, but with various spokes pointing inward. These are visible in Fig \ref{fig:FC1}. The purpose of the spokes is to reduce the accumulation of space charge. As the beam comes in, it will ionize many of the gas-atoms with which it collides. This will create a number of He$^+$--$e^-$ pairs. The electrons move towards the highest potential in the region fast enough as to not be a concern. However, the He$^+$ accumulates fast enough that it produces a space charge effect, pushing incoming ions of interest radially outward and resulting in transport losses \cite{Takamine2005}. The presence of the spokes offers nearby physical obstacles with which the He$^+$ ions can quickly collide and de-ionize, thus mitigating the amount of space charge in that region. 

The largest spokes measure 85 mm in length, and define an interior diameter of 42 mm about the central axis. The medium and small spokes measure 65 mm and 34 mm in length respectively. The inner diameter of the electrodes in sections 1-3, excluding the spokes, is 213 mm. A conservative estimate of beam loss from the spokes, assuming the beam enters the gas catcher uniformly and parallel across the window, yields a maximum of 2\% of the beam lost.

The cone electrodes decrease in diameter along the beam axis in such a way that they define a cone angle of 27$\degree$. The exit nozzle is a de Laval type, with a conical opening and exit that both taper down to a minimum diameter of 1.6 mm

\subsection{Circuit Design}
This gas catcher consists of three independent RF circuits, and two independent DC circuits. The cone and section 1 each have their own RF circuit, while sections 2 and 3 share RF. Meanwhile the entire body (sections 1-3) shares a DC circuit, and the cone has its own. 

In order to optimize RF power delivery to the circuit, the transformer's primary inductor must be impedance matched to the RF generators which supply power. The generators used here are T$\&$C Power Conversion Inc, AG-1020 RF broadband power source, and have an output impedance of 50 $\Omega$. RF signals travel from the generator to an impedance matching transformer. The primary inductor is made of 14 AWG magnet wire. This coil is placed inside the secondary inductor which sends the signal to the rest of the electrodes via copper strips, supplying RF voltage across all the electrodes to which they are connected. The electrodes themselves are each separated from this RF backbone by 1 $\mu F$ capacitors. 

\begin{figure}
    \centering
    \includegraphics[width=\linewidth]{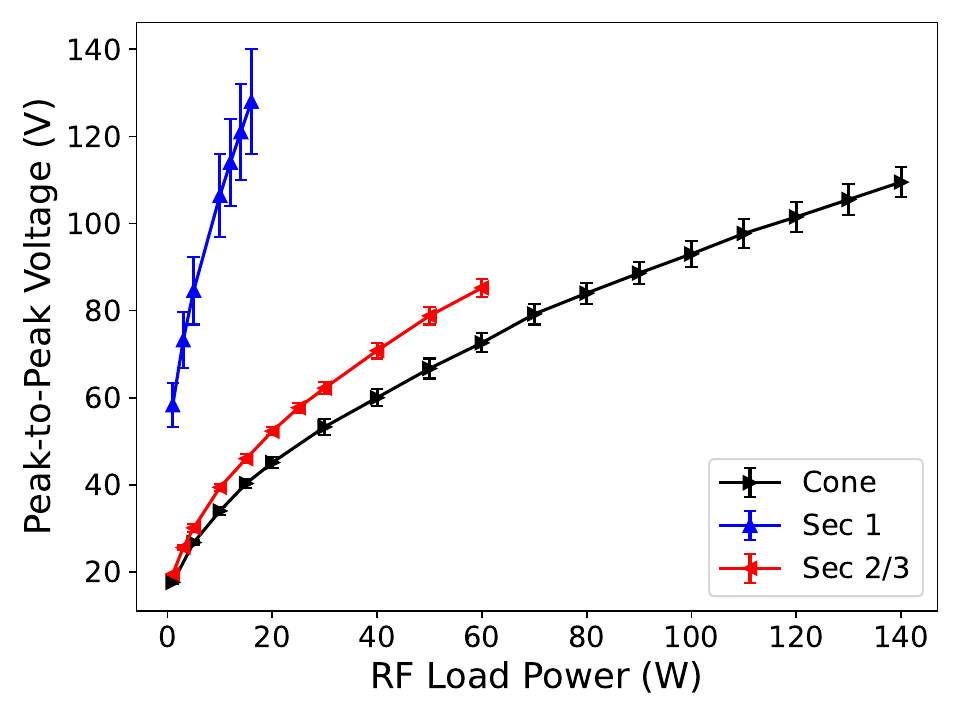}
    \caption{Relationship between applied power and corresponding peak-to-peak voltage for each of the RF sections. Voltage amplitudes were measured at both ends of the secondary pickup coil in order to measure RF phase difference. Uncertainties denote the difference between signals of opposite phase.}
    \label{fig:RFVoltage}
\end{figure}

The number of turns in the primary inductor coil was adjusted so as to supply the appropriate impedance at the circuit's resonant frequency. When the primary and secondary coils are tuned to match the impedance of the RF supply, the resonant frequencies of sections 2/3, 1, and the cone are 2.43 MHz, 2.84 MHz, and 2.47 MHz, respectively. At these resonances no reflected power was observed for the power outputs used in these measurements. Figure \ref{fig:RFVoltage} shows the peak-to-peak voltage, measured on either end of the secondary pickup coil for each of the RF sections, for a given load power supplied by the RF generator.

Running in parallel to the RF backbone, but on the other side of the capacitors, is the DC line, consisting of an alternating series of electrodes and 2.2 $k\Omega$ resistors, such that each electrode sees a small potential drop after each resistor. 

The cone's circuit is on the exterior of the chamber wall, and each electrode has its own electrical feed-through in the chamber wall, while the body sections have the resistor/capacitor chain on the interior. The exception to this is the connection between body sections, which is located on two breadboards on the exterior of the chamber. DC connections between sections 1, 2, and 3 have 90 k$\Omega$ resistors between them, providing a larger potential drop at the end of each segment. In addition, these external connections are used to supply voltage to the chamber walls, so that the walls of each section carry the same voltage as the lowest electrode in that section. This reduces the potential difference between the electrodes and prevents discharges between the walls and the electrodes, allowing for a larger DC drag field to be applied across these sections.

Because the body and cone of the gas catcher maintain two distinct potential gradients, four power supply channels are required to define a high and low voltage for each of them. The lower voltage channel must either be able to sink current, or a resistor to ground must be placed between the last electrode and the power supply. In the case of this commissioning the latter was used, and current-sinking resistors of 33 k$\ohm$ and 9.8 k$\ohm$ were used for the cone-low and body-low electrodes, respectively.

\section{Experimental Setup}
\begin{figure}
    \centering
    \includegraphics[width = 0.75\linewidth]{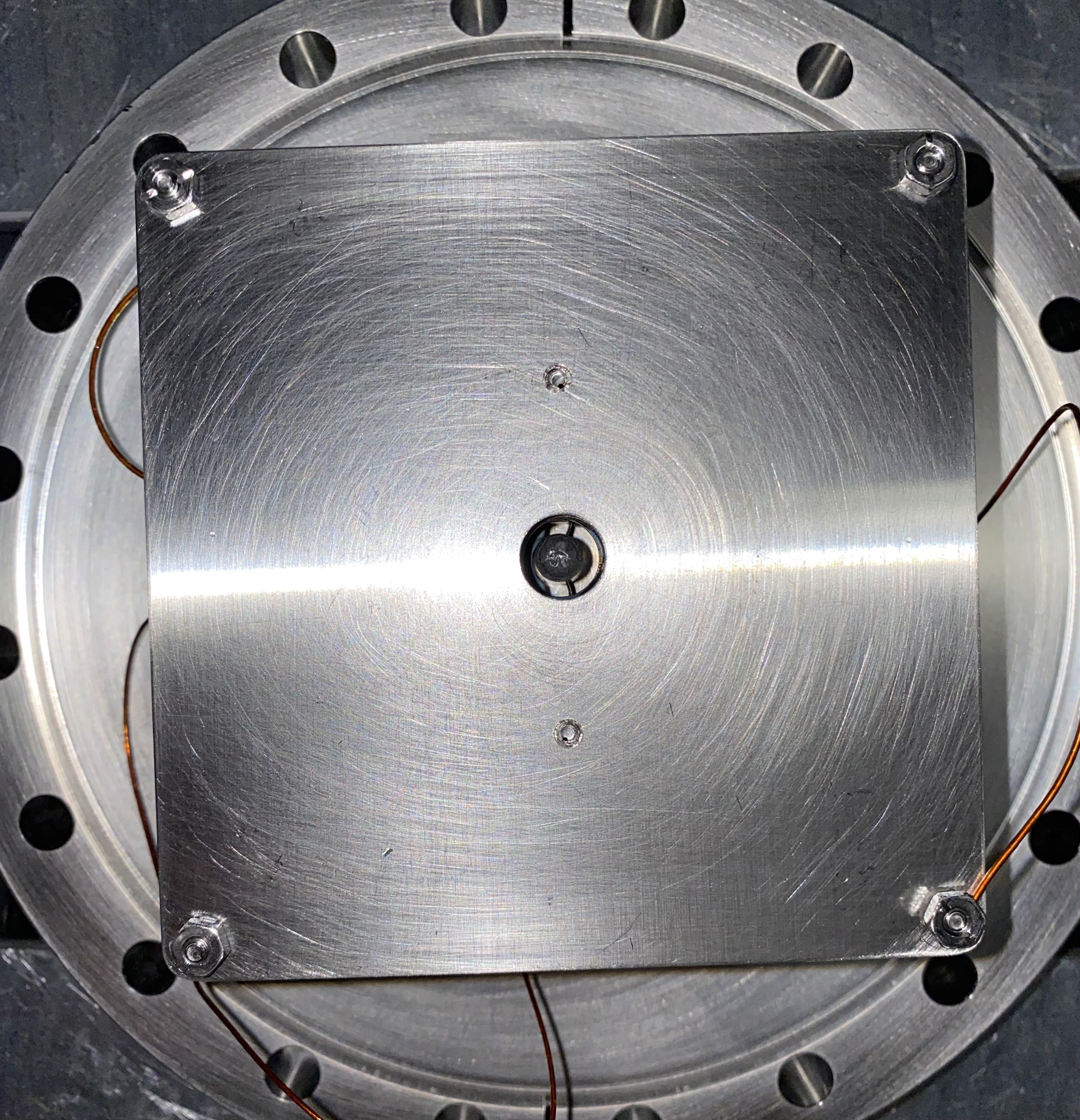}
    \caption{Head-on view of the ion  source. The plate and the source were electrically isolated from each other and could be independently biased.}
    \label{fig:IS_int}
\end{figure}

Offline beam was provided by an ion source mounted on an 8'' Conflat flange. The source mount included a large square anode plate, as shown in Fig. \ref{fig:IS_int}, with a 12 mm opening at the center, behind which the ion source is housed. A potential difference between this anode plate and the source allows for a focused beam to be extracted from the source.

\begin{figure}
    \centering
    \includegraphics[width=0.75\linewidth]{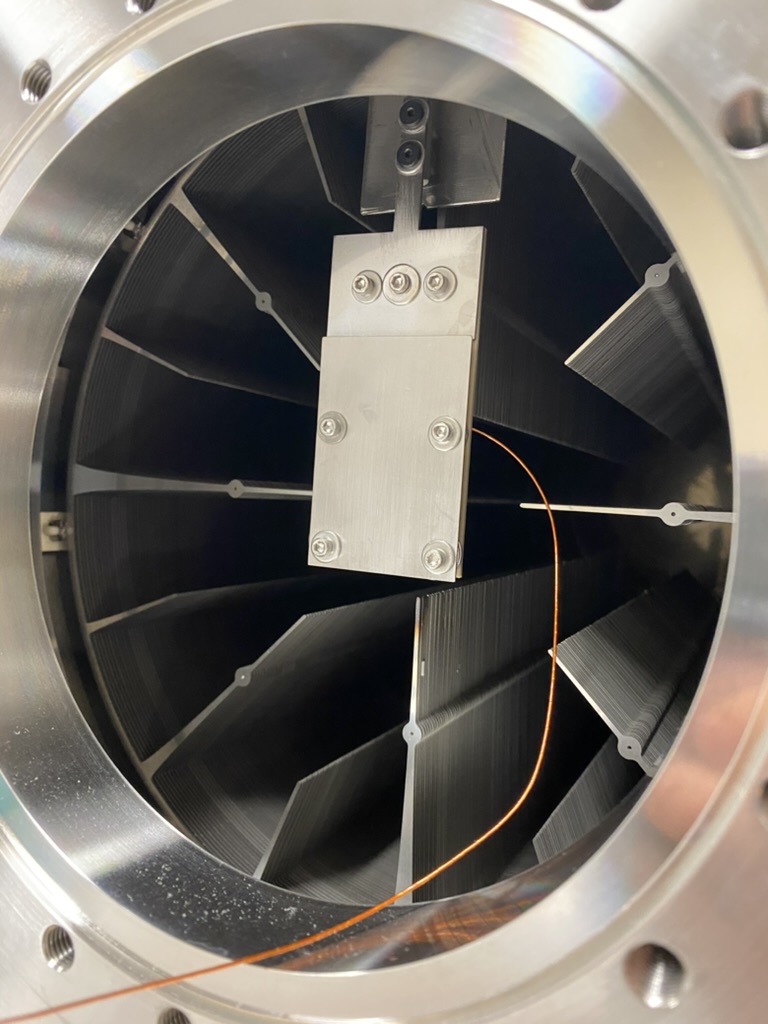}
    \caption{View of the entrance-side collection plate. The plate was attached to a linear drive fed through the top of the chamber. A thin, Kapton insulated copper wire carried current readings from the plate to a BNC feedthrough.}
    \label{fig:FC1}
\end{figure}

The material at the surface of the thermionic source was natural abundance potassium (93\% \isotope{39}{K} and 7\% \isotope{41}{K}). Beam current was measured on two simple, stainless steel collection plates, referred to as FC1 and FC2, though they were not full-fledged Faraday cups as the name might imply. FC1, shown in Fig. \ref{fig:FC1}, was mounted on a linear drive directly in front of the central aperture of the anode plate. FC2 was mounted just downstream of the exit nozzle. Efficiencies were calculated by comparing the current readings on these two plates. 

As such, in this configuration, there were nine locations in the circuit to which a voltage could be directly applied. These are referred to as the source/anode, window, FC1, body high, body low, cone high, cone low, and FC2. Their locations are noted in the diagram shown in Fig. \ref{fig:GCDiagram}. 

One notable effect that needed to be taken into account was that higher pressures have a larger cooling effect on the ion source. This necessitates running a higher heating current through the source in order to produce measurable amounts of current, which reduces its lifetime. For pressures above 66 mbar, typical heating currents ranged from 3.0 -- 3.2 A. 

Beam current measurements were done with a Keithley 6514 electrometer. The connection to FC1 and FC2 was done with a tri-axial cable. This allowed for the electrometer interior to be floated at the same voltage as the collection plate in order to obtain a more precise current reading. Pressure was measured using an MKS Baratron 121A absolute pressure manometer, with a sensitivity of 0.1 mbar. The pressure was manually regulated using a fine adjustment needle valve and a high pressure regulator on the helium gas cylinder. The volume behind the exit nozzle was evacuated using a 55 $m^3$/hr EcoDry 65 vacuum pump. This pumping configuration maintained a pressure on the order of 1-5 mbar in the region of FC2. 

The production rate of the ion source was found to naturally have small, slow variations over time. To account for this, beam production measurements on FC1 were performed before and after each parameter scan, and their timestamps were used to linearly interpolate the produced current at the time of each measurement on FC2. This interpolated current was used to determine the transport efficiency.

\section{Transport Tests}
In order to efficiently stop the \textit{TwinSol} beams, without overloading the downstream differential pumping system, the planned operating pressures of the gas catcher will be between 40-70 mbar. For this reason, the potassium-ion transport in the gas catcher was investigated at three different pressures below, within, and above that range, namely: 33, 66, and 100 mbar. For each of these pressures the potential on each electrodes was varied to maximize the transport efficiency. These optimal settings are presented in Table \ref{tab:nominaldc}, and the optimal RF settings are presented in Table \ref{tab:nominalrf}. In the following section we present the effect on the transport efficiency of varying several key operational parameters. These are the potential difference across the body, $V(\text{body high}) - V(\text{body low})$, between the body and the cone, $V(\text{body low}) - V(\text{cone high})$, across the cone, $V(\text{cone high}) - V(\text{cone low})$, as well as the RF amplitude of the cone and  body sections.

Transport efficiency scans were performed by varying the potential difference across the two electrodes being studied, and scaling every electrode upstream of the electrode farther upstream, such that the relative voltage values are unchanged in the upstream region. Likewise, the scans of RF amplitude were performed with all other parameters held fixed.

\begin{table}[]
    \centering
    \begin{tabularx}{0.7\linewidth}{YY}
        \hline
        Electrode& Voltage (V) \\
        \hline
        \hline
        FC2 & -50 \\
        Cone Low & 0  \\
        Cone High & 25 \\
        Body Low & 50 \\
        Body High & 80 \\ 
        FC1 & 90 \\
        Window & 110 \\
        Anode & 140 \\
        Source & 160 \\
        \hline
    \end{tabularx}
    \caption{Optimized DC voltages for tunable electrodes.}
    \label{tab:nominaldc}
\end{table}

\begin{table}[]
    \centering
    \begin{tabularx}{0.75\linewidth}{YYYY}
        \hline
        Section & Power (W) & Voltage Amplitude (V) & Frequency (MHz) \\
        \hline
        \hline
        Cone & 40 & 30 & 2.479 \\
        1 & 10 & 53 & 2.840 \\
        2/3 & 50 & 40 & 2.112 \\
        \hline
    \end{tabularx}
    \caption{Nominal operating RF power parameters.}
    \label{tab:nominalrf}
\end{table}

\subsection{Production and measurement of ion beam current}
These measurements were performed over the course of several days. At the end of each day the gas catcher was evacuated overnight using a 70 L/s turbo molecular pump. Base pressures observed on the following day were typically around $5\times 10^{-7}$ torr. Each day the ion source was warmed up for approximately two hours, with helium flowing at the desired pressure, until a stable pressure and ion current on FC1 were observed. Current readings on FC1 were then taken every 10-20 minutes to monitor ion beam production throughout the testing process. For all measurements under study, a constant potential difference of 20 V was maintained between the ion source and anode plate. Table \ref{tab:sourceheat} shows typical operational parameters for the ion source at the studied pressures.

\begin{table}[]
    \centering
    \begin{tabularx}{0.7\linewidth}{YYYY}
        \hline
        Pressure (mbar) & $I_{\text{heat}}$ (A) & $V_{\text{heat}}$ (V) & $I_{\text{beam}}$ (pA)\\
        \hline
        \hline
        33  & 3.08 & 7.0 & 45\\
        66  & 3.20 & 7.2 & 45\\
        100 & 3.26 & 7.3 & 80\\
        \hline
    \end{tabularx}
    \caption{Typical voltage and and current applied to the potassium ion source at the three pressures studied. The quoted beam production current is included as a representative value, and does not indicate a constant current value across measurements.}
    \label{tab:sourceheat}
\end{table}

\begin{figure}
    \centering
    \includegraphics[width=\linewidth]{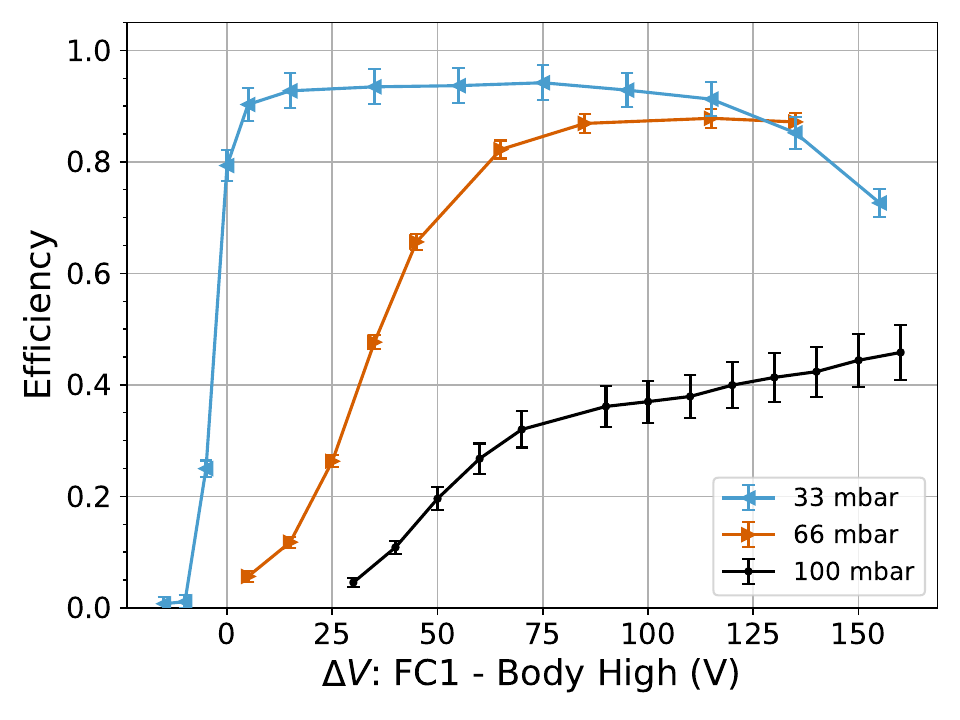}
    \caption{Transport efficiency versus voltage difference between FC1 and body high electrode at the three pressures studied.}
    \label{fig:DCFC1}
\end{figure}

After each current reading with FC1, this electrode was raised out of the beam path using a linear motion actuator. While FC1 was retracted, its influence on the ion motion was still observed. Hence, at each pressure the potential of electrodes in the vicinity of the ion source were adjusted so as to maximize the ion transport efficiency. These electrodes include the source/anode pair, the window (the chamber wall surrounding the ion source and anode), and FC1. The result of one of these scans is shown in Fig. \ref{fig:DCFC1}, where the 33 mbar curve shows that outside a certain range, the voltage on FC1 steers the beam too far in one direction or the other.

\subsection{Transport to the cone}
\begin{figure}
    \centering
    \includegraphics[width=\linewidth]{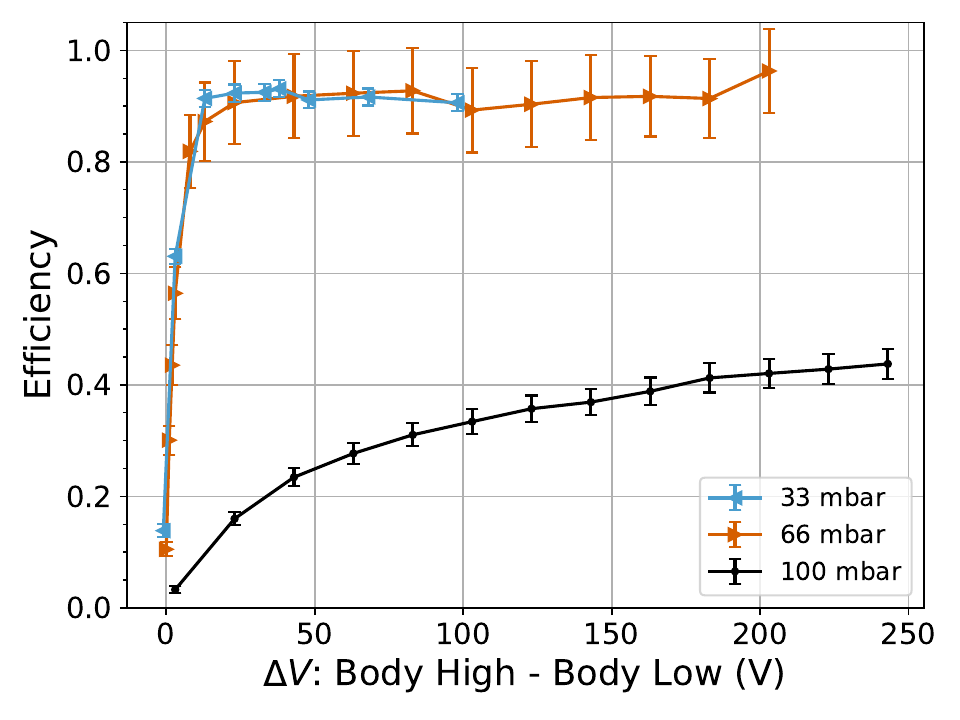}
    \caption{ Transport efficiency vs DC gradient across the body region (sections 1-3) at the three pressures studied.}
    \label{fig:DCBH}
\end{figure}

Upon leaving the ion source region, the beam enters the main body of the gas catcher, where the transport quality is dominated by RF amplitude and the strength of the drag field. As indicated in Fig. \ref{fig:DCBH}, at pressures of 33 or 66 mbar, only a small potential difference across the body (10-15 V) is necessary to fully transport the ions through that section. On the other hand, at 100 mbar, even a potential difference of 243 V was insufficient to efficiently transport the ions. Though the trend suggests that increasing the voltage would have continued to increase transport efficiency, electrical discharges were observed for $\Delta V >$ 250 V. For this reason, various tests at 100 mbar were unable to find a true optimum value. 

\begin{figure}
    \centering
    \includegraphics[width=\linewidth]{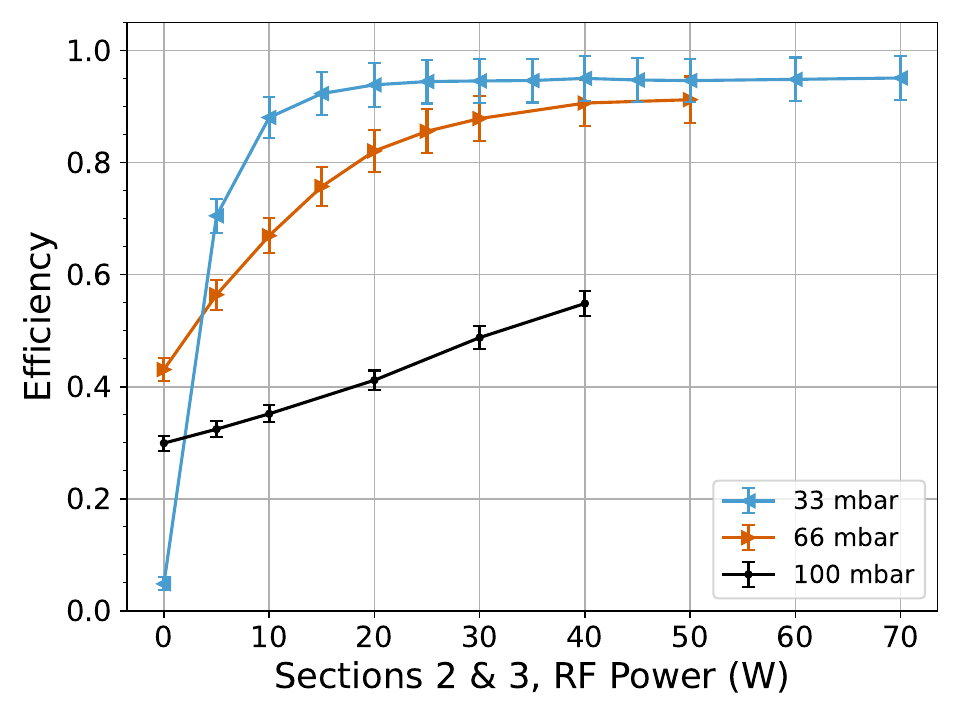}
    \caption{Transport eficiency vs RF power in sections 2 and 3 at the three pressures studied.}
    \label{fig:RF23}
\end{figure}

Figure \ref{fig:RF23} shows the containment effect of Section 2/3 RF on the transported ion, clearly demonstrating that some RF is required for sections 2 and 3, regardless of the pressure studied. As the pressure increased, a larger repelling force (in the form of a larger RF amplitude) was required to keep the ions away from the electrodes. Sufficient repulsion was achieved at both 33 and 66 mbar. Unfortunately, this was not the case at 100 mbar as a result of the discharge limitations previously mentioned. 

The interesting behavior at low RF amplitude, where the efficiency drops more quickly at lower pressure, is most likely due to the different drag fields used at each of these pressures. A potential difference of only 38 V was used at 33 mbar while 143 V and 193 V were used at 66 and 100 mbar, respectively. The 33 mbar scan was not repeated with a stronger drag field since the purpose of these RF scans was simply to demonstrate the plateauing behavior of transport efficiency vs RF repulsion.

\begin{figure}
    \centering
    \includegraphics[width=\linewidth]{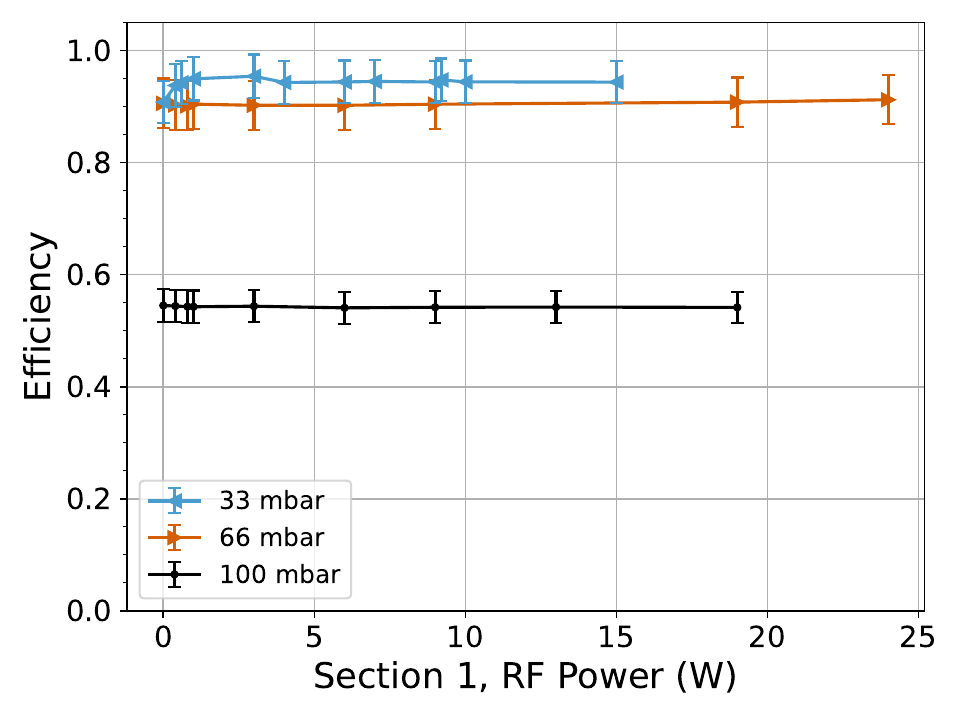}
    \caption{Transport efficiency vs RF power on section 1 at the three pressures studied.}
    \label{fig:RFsec1}
\end{figure}

The absence of spokes in section 1, and consequently the much larger distance of the electrodes from the ion path, meant that the amplitude of the RF in section 1 did not have any effect on the ion beam transport, as is evident in Fig. \ref{fig:RFsec1}.

\begin{figure}
    \centering
    \includegraphics[width=\linewidth]{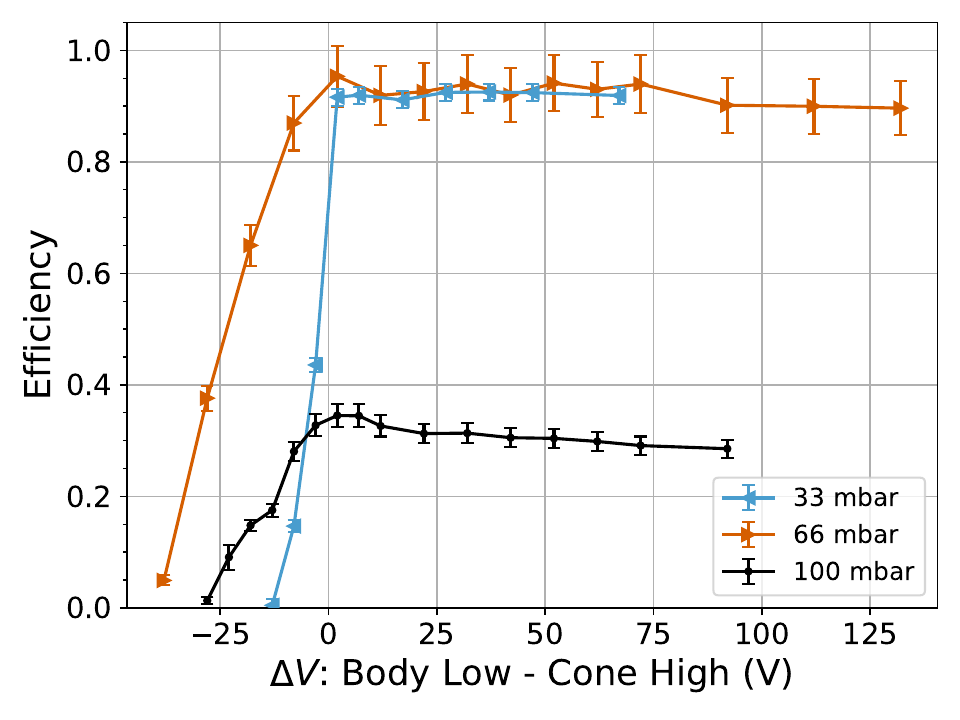}
    \caption{Transport efficiency vs voltage difference between body and cone regions at the three pressures studied.}
    \label{fig:DCBL}
\end{figure}

Figure \ref{fig:DCBL} shows the effect of the potential difference between the body electrode farthest downstream (body low) and the most upstream cone electrode (cone high). For all pressures studied, a small potential difference, of around 2 V, is sufficient to efficiently  transport the ions between the two sections. The transport efficiency at 33 and 66 mbar flattens at higher potential differences, while at 100 mbar it slowly decreases.

\subsection{Extraction Through the Nozzle}

\begin{figure}
    \centering
    \includegraphics[width=\linewidth]{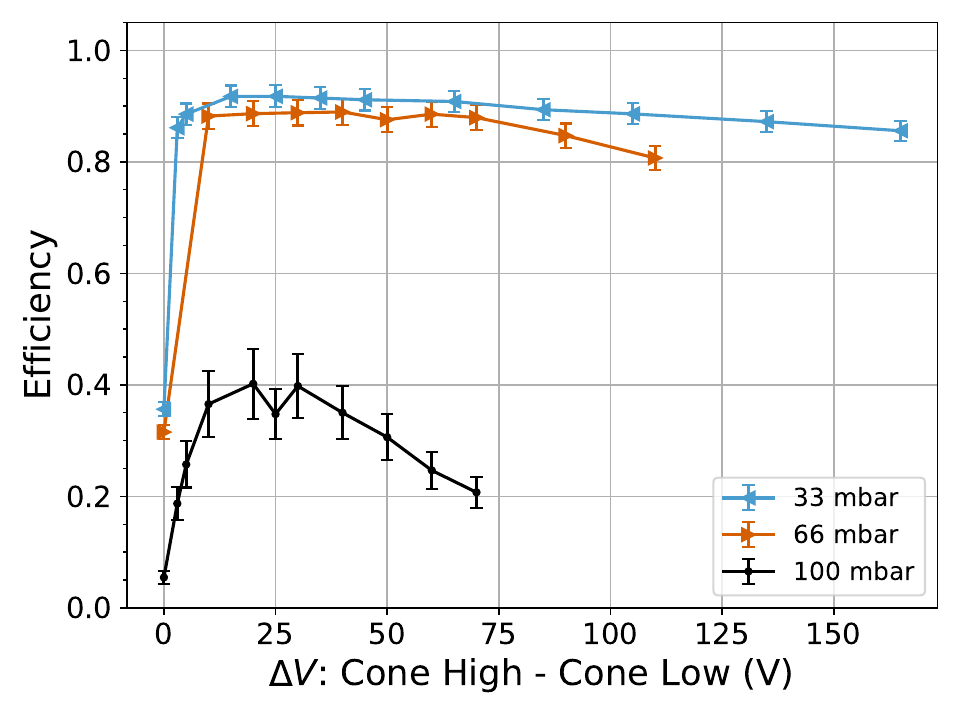}
    \caption{Transport efficiency vs DC gradient across the cone at the three pressures studied. }
    \label{fig:DCCH}
\end{figure}

\begin{figure}
    \centering
    \includegraphics[width=\linewidth]{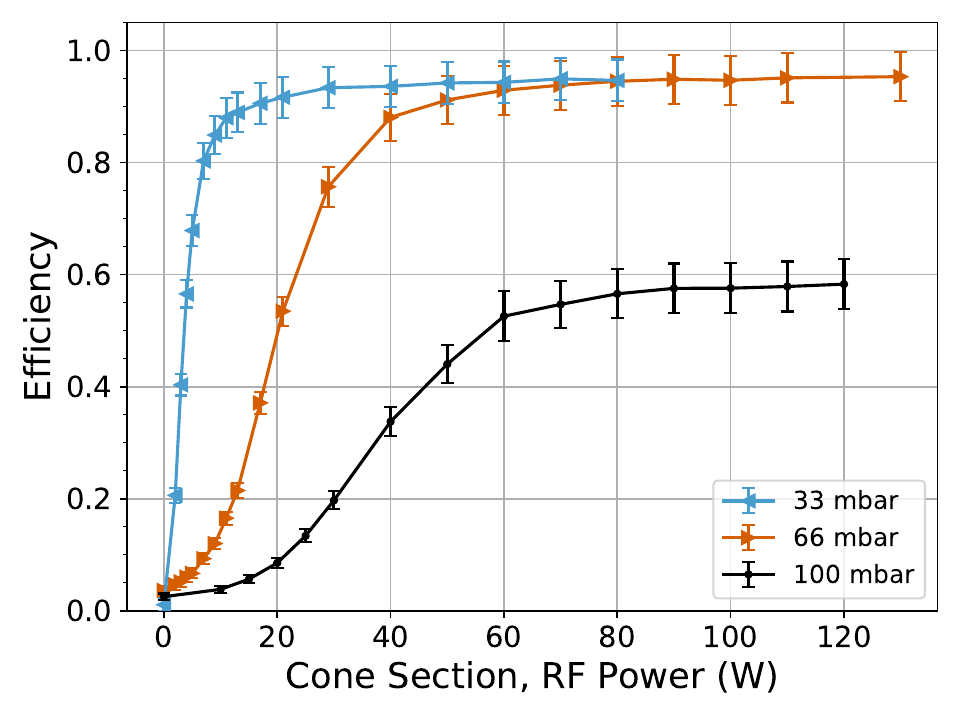}
    \caption{Transport efficiency vs RF power on cone electrodes at the three pressures studied. Here the expected plateau behavior is clearly visible.}
    \label{fig:RFCone}
\end{figure}

The cone of the gas catcher is made of a series of concentric, stacked rings of decreasing diameter that form a funnel, leading the ions to the nozzle. Due to the angular slant the electric field has towards the cone electrodes, the transport efficiency in the cone section will be more sensitive to the potential difference across that section as compared to the body. This is shown in Fig. \ref{fig:DCCH} where, for all three pressures, an optimum drag field is observed. The lower the pressure, the wider the range of potential differences leading to efficient transport. Figure \ref{fig:RFCone} shows that at all pressures there is a saturation in transport efficiency, where the minimum RF amplitude for saturation increases with pressure. This indicates that most of the ion losses at 100 mbar do not occur in the cone section but rather in the body.

It was observed that a potential difference of at least 10 V was necessary between the nozzle and FC2 for efficient and stable extraction of the ion beam. At the same time, for all pressures studied, the most efficient extraction was observed when the most downstream electrode of the cone was at the same potential as the nozzle. Hence, for all measurements, that electrode was grounded, i.e. $V(\text{cone low}) = 0$ V.

\section{Conclusion}
The St. Benedict facility, in the Nuclear Science Laboratory at the University of Notre Dame, aims to measure the beta-neutrino angular correlation parameter in superallowed mixed beta-decay transitions between mirror nuclei. The first component of this facility is a large volume gas catcher which will thermalize the 10 - 40 MeV radioactive ion beams from the \textit{TwinSol} facility via collision with a helium gas at pressures around 50 mbar. The gas catcher has been commissioned off-line using a thermionic source of potassium ions, placed inside the gas volume at the location of the entrance window. Ions were transported through the entire device to a collection plate located after the nozzle, demonstrating the efficacy of the gas catcher in transporting and extracting thermalized ions. Transport efficiencies above 95\% were observed at pressures of 33 and 66 mbar, which are around the planned operating pressure. Efficient transport at the higher pressure of 100 mbar was found to require a larger drag field and RF amplitude in the body section of the catcher. These results represent a characterization of the capacity of the gas catcher to transport ions, decoupled from its ability to stop beam. On-line operation may introduce effects that impact the behavior of the trends observed in this study; however, these data offer insight into the behavior of the gas catcher which will facilitate on-line characterization. With this successful off-line commissioning, the gas catcher is now ready for on-line operation.

\section{Acknowledgments}
This work was carried out with the support of the National Science Foundation under grants PHY-1725711, PHY-2310059, PHY-2050527, the University of Notre Dame, as well as the US Department of Energy, Office of Science, Office of Nuclear Physics under Contract No. DEAC02-06CH11357 (ANL).

\bibliographystyle{elsarticle-num} 
\bibliography{library}

\end{document}